\documentclass[aps,reprint,amsmath,amssymb,superscriptaddress,showpacs,prb]{revtex4-1}
\usepackage{siunitx}

\usepackage{graphicx}

\newcommand{\ve}{\varepsilon}
\renewcommand{\alpha}{ a }

\begin{document}

\title{Fundamental characteristic length scale for the field dependence of hopping charge transport
in disordered organic semiconductors}

\author{A.~V.~Nenashev}
\affiliation{Institute of Semiconductor Physics, 630090 Novosibirsk, Russia}
\affiliation{Department of Physics, Novosibirsk State University, 630090 Novosibirsk, Russia}

\author {J.~O.~Oelerich}
\affiliation{Department of Physics and Material Sciences Center,
Philipps-University, D-35032 Marburg, Germany}

\author{A.~V.~Dvurechenskii}
\affiliation{Institute of Semiconductor Physics, 630090 Novosibirsk, Russia}
\affiliation{Department of Physics, Novosibirsk State University, 630090 Novosibirsk, Russia}

\author{F.~Gebhard}
\affiliation{Department of Physics and Material Sciences Center,
Philipps-University, D-35032 Marburg, Germany}

\author {S.~D.~Baranovskii}
\affiliation{Department of Physics and Material Sciences Center,
Philipps-University, D-35032 Marburg, Germany}

\date{\today}

\begin{abstract}
	Using analytical arguments and computer simulations we show that the dependence of the hopping carrier mobility on the electric field $\mu(F)/\mu(0)$ in a system of random sites is determined by the localization length $a$, and not by the concentration of sites $N$. This result is in drastic contrast to what is usually assumed in the literature for a theoretical description of experimental data and for device modeling, where $N^{-1/3}$ is considered as the decisive length scale for $\mu(F)$. We show that although the limiting value $\mu(F \rightarrow 0)$ is determined by the ratio $N^{-1/3}/a$, the dependence $\mu(F)/\mu(0)$ is sensitive to the magnitude of $a$, and not to $N^{-1/3}$. Furthermore, our numerical and analytical results prove that the effective temperature responsible for the combined effect of the electric field $F$ and the real temperature $T$ on the hopping transport via spatially random sites can contain the electric field only in the combination $eFa$.
\end{abstract}


\maketitle

\section{Introduction}

Organic semiconductors attract currently much attention in the scientific community as materials desired for applications in modern electronics. The term “organic semiconductors” covers a large class of materials with a broad variety of properties. Organic semiconductors can be fabricated in crystalline form, as for instance, pentacene and ruberene \cite{Ostroverkhova2016}. The energy spectrum in such materials has a classical band structure with charge carriers moving as free particles or polarons in the conduction and valence bands. The main focus in research on organic materials is put, however, on organic disordered semiconductors (ODSs), such as polymers and low-molecular-weight systems \cite{Silinsh1970,Bassler1993,Bassler1981,Borsenberger1993,Auweraer1994,Pope1999,Hadziioannou2000,Brabec2003,Bruetting2005,Baranovski2006,Schwoerer2007,Sun2008,Tessler2009,Meller2010,Geoghegan_2013,Baranovskii2014,Kuik2014,Anna_Heinz_2015,Nenashev_Topical_2015,Laquai2015}. The interest in ODSs is caused by their optoelectronic features and by easy manufacturing, as compared to organic crystals. In contrary to crystalline materials, ODSs possess neither structural regularity, nor spatially extended electronic states. Instead, electronic states in ODSs are spatially localized \cite{Silinsh1970,Bassler1981,Bassler1993,Borsenberger1993,Auweraer1994,Pope1999,Hadziioannou2000,Brabec2003,Bruetting2005,Baranovski2006,Schwoerer2007,Sun2008,Tessler2009,Meller2010,Geoghegan_2013,Baranovskii2014,Kuik2014,Anna_Heinz_2015,Nenashev_Topical_2015,Laquai2015}.  This happens because the overlap integrals for the weak Van-der-Waals interactions between neighboring structural units  (molecules or molecular complexes) in ODSs are much smaller than the energy scale of disorder, which prevents the formation of extended electronic states \cite{Silinsh1970,Bassler1981,Bassler1993}.   Therefore, charge transport in ODSs is due to incoherent tunneling (hopping) of charge carriers between localized states that are randomly distributed in space \cite{Silinsh1970,Bassler1981,Bassler1993,Borsenberger1993,Auweraer1994,Pope1999,Hadziioannou2000,Brabec2003,Bruetting2005,Baranovski2006,Schwoerer2007,Sun2008,Tessler2009,Meller2010,Geoghegan_2013,Baranovskii2014,Kuik2014,Anna_Heinz_2015,Nenashev_Topical_2015,Laquai2015}. Our paper deals with the description of charge transport in this hopping regime and the results are valid for ODSs, not for organic crystals.

The most popular theoretical model to describe charge transport in ODSs is the so-called Gaussian Disorder Model (GDM), according to which localized states have a Gaussian energy distribution\cite{Silinsh1970,Bassler1981,Bassler1993,Oelerich2012}
\begin{align}
	\label{DOS_Gauss}
	g(\varepsilon) = \frac{N}{\sigma\sqrt{2\pi}}
	\exp\left(-\frac{\varepsilon
	^{2}}{2\sigma^{2}}\right) \, .
\end{align}
Here, $\sigma$ is the energy scale of the spectrum, usually
estimated\cite{Bassler1993} in ODSs to the order of $\sigma \approx 0.1$ eV
 and $N$ is the concentration of randomly distributed
localized states (sites). A typical estimate\cite{Baranovski2006,Baranovskii2014} for the latter parameter is
between $N \simeq 10^{20}$ cm$^{-3}$ and $N \simeq 10^{21}$ cm$^{-3}$.

The hopping rates are usually assumed \cite{Bassler1993} to be described by the Miller-Abrahams expression\cite{Miller1960}. For each pair of sites $(i,j)$, the rate $\nu_{ij}$ is determined by their energy difference $\ve_j-\ve_i$ and position difference $\mathbf{r}_{ij} \equiv \mathbf{r}_j-\mathbf{r}_i$:
\begin{equation}
	\label{eq:nu-ij}
	\nu_{ij} = \nu_0 \exp\left( -\frac{2|\mathbf{r}_{ij}|}{\alpha} \right) \gamma(\ve_j-\ve_i+e\mathbf{F}\cdot\mathbf{r}_{ij})
\end{equation}
with
\begin{equation}
	\label{eq:gamma}
	\gamma(\Delta\ve) =
	\begin{cases}
		\exp(-\Delta\ve / kT) , & \text{if $\Delta\ve>0$,} \\
		1,                      & \text{otherwise}  ,
	\end{cases}
\end{equation}
where $\alpha$ is the localization length of charge carriers, $\mathbf{F}$ is the electric field, and $\nu_0$ is a prefactor determined by the tunneling mechanism. The localization length $\alpha$ in ODSs is estimated \cite{Gill1972,Rubel2004} at the order of $10^{-8}$ cm, which is much smaller than the intersite distance $N^{-1/3}$. Therefore, we follow the usual assumption \cite{Silinsh1970,Bassler1981,Bassler1993,Borsenberger1993,Auweraer1994,Pope1999,Hadziioannou2000,Brabec2003,Bruetting2005,Baranovski2006,Schwoerer2007,Sun2008,Tessler2009,Meller2010,Geoghegan_2013,Baranovskii2014,Kuik2014,Anna_Heinz_2015,Nenashev_Topical_2015,Laquai2015} that $\alpha$ can be considered to be independent of the concentration of sites $N$.

While powerful and transparent analytical theoretical tools have been developed to describe the dependencies of the hopping mobility $\mu$ on $T$, $N$, $\alpha$, $\sigma$, and on the concentration of carriers $n$, as highlighted in recent reviews \cite{Baranovski2006,Tessler2009,Baranovskii2014,Nenashev_Topical_2015,Ostroverkhova2016}, theoretical studies of the dependence $\mu(F)$ have mostly been focused on computer simulations. The group of B\"{a}ssler simulated $\mu(F)$ on a cubic lattice and fitted results in the form of the parameterized equation \cite{Borsenberger1991,Bassler1993,Anna_Heinz_2015}
\begin{align}
	\label{eq:ElectrFieldBassler}
	\begin{split}
	\mu(F)= & \mu_{0} \exp{\left\{-\left(\frac{2}{3}
	\frac{\sigma}{kT}\right)^{2}\right\}} \\
	        & \times \exp\left\{\widetilde{C}\left[\left(\frac{\sigma}{kT}\right)^{2}
	-B\right]\sqrt{F}\right\} \, ,
	\end{split}
\end{align}
where $\mu_0$ is a field-independent prefactor.

Two parameters, $\widetilde{C}$ and $B$ are involved in this fitting. The parameter $\widetilde{C}$ is assumed to depend on the lattice constant $b$ (distance between localization sites) having the value  $\widetilde{C} = 2.9 \times 10^{-4}$ cm$^{1/2}$/V$^{1/2}$ for $b=0.6$ nm \cite{Borsenberger1991,Bassler1993,Anna_Heinz_2015}. Although simulations were performed on regular cubic grids, a non-diagonal disorder has been introduced into simulations by B\"{a}ssler \textit{et al.} \cite{Borsenberger1991,Bassler1993,Anna_Heinz_2015}  in order to mimic spatial disorder. The exponent $(2|\mathbf{r}_{ij}|/a)$ in Eq. (\ref{eq:nu-ij}) was rewritten in the form $2\lambda b|\mathbf{r}_{ij}|/b$ , where $b$ is the lattice spacing, and the parameter $\lambda$ can be viewed as the inverse localization length. The factor $2 \lambda b$ was distributed in a Gaussian manner with the width $\Sigma$ around the value $2 \lambda b=10$. The parameter $B$ in Eq.~(\ref{eq:ElectrFieldBassler}) was set equal to $B = 2.25$ for $\Sigma < 1.5$ and to $B = \Sigma^{2}$ for $\Sigma \geq 1.5$.  Equation~(\ref{eq:ElectrFieldBassler}) is one of the most frequently used equations in the context of organic
semiconductors \cite{Tessler2009,Baranovskii2014}.


A similar approach to determine $\mu(F)$ was used by Pasveer \textit{et al}. \cite{Pasveer2005}, who reduced the lattice GDM of B\"{a}ssler \textit{et al}. to the case $\Sigma = 0$ and herewith completely eliminated spatial disorder. Calculating numerically $\mu(F)$  in the framework of this reduced GDM on a cubic lattice, Pasveer \textit{et al}. fitted results to the analytical formula
\begin{align}
	\label{eq:ElectrFieldPasveer_1}
	\mu(T,n,F)\approx\mu(T,n) \phi(T,F) \,
\end{align}
with $\phi(T,F)$ in the form
\begin{align}
	\label{eq:ElectrFieldPasveer}
	\begin{split}
	\phi(T,F) = \exp\Bigg\{
	  & 0.44 \left[\left(\frac{\sigma}{kT}\right)^{3/2}-2.2\right]            \\
	  & \times \left[\sqrt{1+0.8\left(\frac{Feb}{\sigma}\right)^{2}}-1\right]
	\Bigg\} \, ,
	\end{split}
\end{align}
where $b$ is the lattice constant. The latter equations are sometimes considered universal and they are the basis \cite{Coehoorn_Bobbert_Feature2012} for the commercially
available OLED simulation software tools [Simulation
software SETFOS3.2, product of Fluxim (www.fluxim.com);
Simulation software SimOLED3.x, product of Sim4tec
(www.sim4tec.com)].

Pasveer \textit{et al}. \cite{Pasveer2005} mentioned that Eq.(\ref{eq:ElectrFieldPasveer})  ``should merely be considered as a description of the numerical data in a limited parameter range" promising to rationalize this parametrization in future work. We show below that neither Eq.~(\ref{eq:ElectrFieldBassler}) nor Eq.~(\ref{eq:ElectrFieldPasveer}) can be rationalized because they do not contain decisive parameters responsible for the field-dependent mobility $\mu(F)$. Equations (\ref{eq:ElectrFieldBassler}) and (\ref{eq:ElectrFieldPasveer}), which are used by thousands of researchers, were obtained by fitting the numerically simulated data under the assumption that the decisive parameter for the dependence $\mu(F)$ is the intersite distance, parameter $b$ in Eq.~(\ref{eq:ElectrFieldPasveer}). We rigorously prove below that this assumption is wrong and the intersite distance is irrelevant for the field-dependent mobility in disordered systems. One should instead use the localization length $a$ as the decisive length scale determining the field dependence of the hopping carrier mobility  $\mu(F)$. A theoretical recipe on how to describe the dependence $\mu(F)$ in disordered materials will be formulated below, which should encourage researchers to reanalyze their data on $\mu(F)$ in disordered organic semiconductors.

The paper is organized as follows. In Sec. \ref{sec:alpha is relevant on lattices}, we first stay for simplicity in the framework of the reduced GDM used by Pasveer \textit{et al}. \cite{Pasveer2005}, i.e., on a cubic lattice without spatial disorder. We show that already in this oversimplified case, Eq.~(\ref{eq:ElectrFieldBassler}) and Eq.~(\ref{eq:ElectrFieldPasveer}) are incompatible with each other even if the same material parameters in these equations are used. We further show that the results of computer simulations by Pasveer \textit{et al}. \cite{Pasveer2005} are incompatible with the results of computer simulations by B\"{a}ssler \textit{et al}. \cite{Borsenberger1991,Bassler1993,Anna_Heinz_2015}  carried out in the framework of the same reduced GDM on the cubic lattice (i.e. for $\Sigma=0$). Performing our own computer simulations, we prove that the localization length $a$, not even present in Eqs. (\ref{eq:ElectrFieldBassler}) and (\ref{eq:ElectrFieldPasveer}), is responsible for this discrepancy in the simulations and that the localization length affects decisively the field dependence of carrier mobility.

In Sec.~\ref{sec:Teff simulations}, we consider the GDM on spatially random sites, i.e., not anymore on a lattice, and show by computer simulations that the localization length $a$ is the \textit{only} spatial scale responsible for the field—-dependent hopping mobility. Our computer simulations show herewith that the intersite distance, present in the form of lattice constant $b$ in Eq.~(\ref{eq:ElectrFieldPasveer}), is irrelevant for the field—-dependent mobility  $\mu(F)$.

In Sec.~\ref{sec:Teff}, we further show by computer simulations that the dependence of the carrier mobility on the electric field $F$ can be described by inserting the field-—dependent effective temperature $T_{\text{eff}}(F,T)$, instead of the real temperature $T$, into the temperature dependence of the hopping mobility, which is well understood and described at low electric fields \cite{Baranovski2006,Baranovskii2014,Nenashev_Topical_2015}.  Herewith our computer simulations on random sites rigorously prove the idea by Shklovskii \textit{et al}. \cite{Shklovskii1973,Shklovskii1990Fritzsche,Marianer1992,Hess1993,Cleve1995,Jansson2008PRB}, who already suggested many years ago that the field—-dependent effective temperature, which contains the localization length $a$ as the only relevant spatial parameter, describes the combined effects of electric field and temperature on the hopping mobility.

In Sec. \ref{sec:Teff analytical}, we prove the concept of the effective temperature for spatially random sites by analytical calculations.  It is shown that the effective temperature does exist and that it depends on the localization length $a$, and not on the concentration of sites $N$.

Concluding remarks are gathered in Sec.~\ref{sec:conclusions}.

A short version of this work has been made publicly available in Ref. \cite{Nenashev_ARKHIV2017}.


\section{Localization length affects $\mu(F)$ in the lattice model}
\label{sec:alpha is relevant on lattices}

\begin{figure}
	\includegraphics[width=\linewidth]{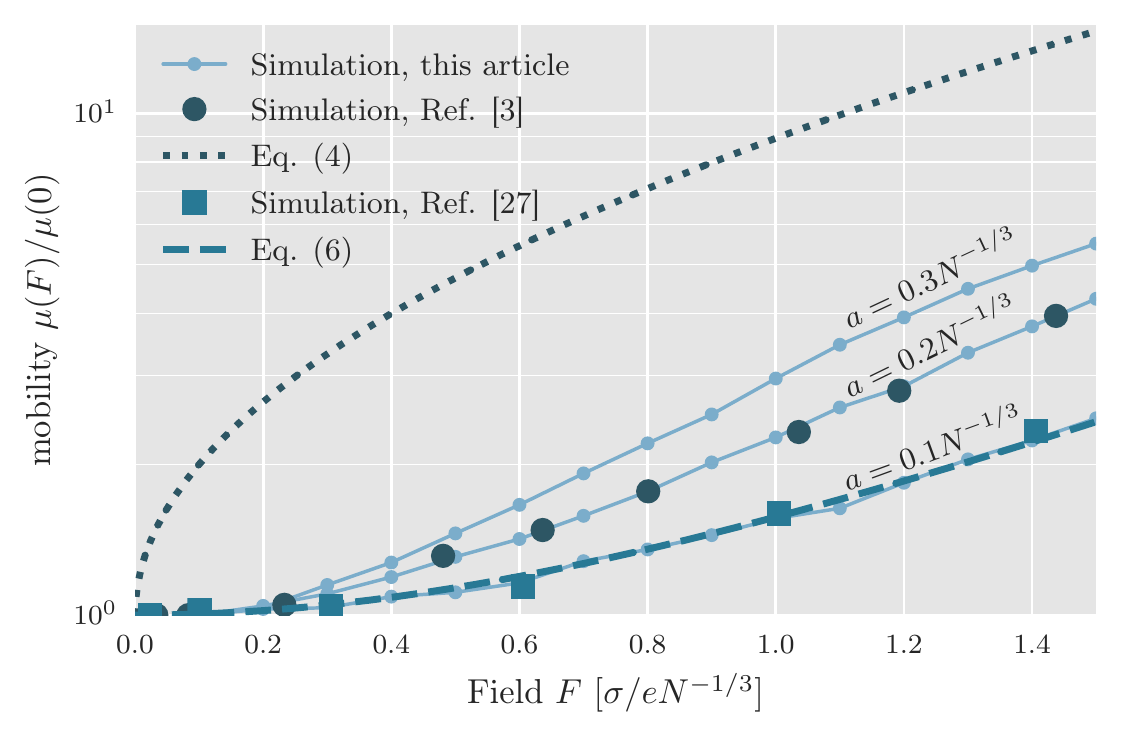}
	\caption{Normalized mobility $\mu(F)/\mu(0)$ for the lattice model calculated via Eqs.~(\ref{eq:ElectrFieldBassler}) and (\ref{eq:ElectrFieldPasveer}) and simulated for $\sigma/kT =3$.}
	\label{fig:Baessler_vs_Pasveer}
\end{figure}

Before considering a realistic case of a spatially disordered system, let us analyze the simulated data on the field--dependent mobility $\mu(F)$ available in the literature \cite{Bassler1993,Pasveer2005}, which were obtained on regular cubic lattices and served for parametrizations by Eqs.~(\ref{eq:ElectrFieldBassler}) and (\ref{eq:ElectrFieldPasveer}). The concentration of sites $N={b^{-3}}$ is used for the plots in Fig.~\ref{fig:Baessler_vs_Pasveer} in order to consist with the data in other figures calculated for random sites.

Let us first check the compatibility of Eqs.~(\ref{eq:ElectrFieldBassler}) and (\ref{eq:ElectrFieldPasveer}) with each other. In order to enable the comparison, we plot the data of B\"{a}ssler \textit{et al}. \cite{Borsenberger1991,Bassler1993} for the case $\Sigma = 0$, i.e., with $B=2.25$, since Pasveer \textit{et al.}\cite{Pasveer2005} simulated for $\Sigma = 0$. The value $T=\SI{300}{\kelvin}$ was used in simulations by B\"{a}ssler \textit{et al}., which gives $\sigma=\SI{0.075}{\electronvolt}$ for $\sigma/kT = 3$. Using the realistic value $\sigma/kT = 3$, we plot by a dotted line in Fig.~\ref{fig:Baessler_vs_Pasveer}  the curve for $\mu(F)/\mu(0)$ given by Eq.~(\ref{eq:ElectrFieldBassler}) and by a dashed line the curve for the function $\phi(T,F)$ given by Eq.~(\ref{eq:ElectrFieldPasveer}). The difference in the dependencies $\mu(F)$ given by Eqs.~(\ref{eq:ElectrFieldBassler}) and (\ref{eq:ElectrFieldPasveer}) for the same sets of parameters is striking.

Trying to reveal the reason for such a large discrepancy, we also plot in Fig.~\ref{fig:Baessler_vs_Pasveer} the simulated data \cite{Bassler1993,Pasveer2005} that served as the basis for fittings by Eqs.~(\ref{eq:ElectrFieldBassler}) and (\ref{eq:ElectrFieldPasveer}). The apparent inability of Eq.~(\ref{eq:ElectrFieldBassler}) to fit the simulated data evidences the poor accuracy of this equation but it can hardly be considered as an issue of fundamental importance. However, it is surely an issue of fundamental importance to elucidate the
difference in the results of the two simulations \cite{Bassler1993,Pasveer2005} for the same value $\sigma/kT = 3$ because the difference between the data obtained in simulations by the group of B\"{a}ssler\cite{Borsenberger1991,Bassler1993} and by Pasveer \textit{et al.}\cite{Pasveer2005}  is comparable to the total effect of $F$ on $\mu$.

The apparent difference in the simulated systems lies in the choice of the parameter $b/\alpha$. While the group of B\"{a}ssler simulated for $b/\alpha=5$, Pasveer \textit{et al.} simulated for $b/\alpha=10$. In order to check the validity of those previous simulations, we carried out simulations on a cubic lattice similar to those carried out by B\"{a}ssler \textit{et al.} and by Pasveer \textit{et al.}. Our data, plotted in Fig.~\ref{fig:Baessler_vs_Pasveer} for $b/\alpha=10; 5; 3$, confirm the data by Pasveer \textit{et al.} with $b/\alpha=10$ and the data by B\"{a}ssler \textit{et al.} with $b/\alpha=5$, implying that the computer simulations by both research groups \cite{Borsenberger1991,Bassler1993,Pasveer2005} were correct. However, it has not been recognized in previous simulations that the shape of the dependence $\mu(F)/\mu(0)$ is sensitive to the choice of $b/\alpha$.

This result shows that neither Eq.~(\ref{eq:ElectrFieldBassler}) nor Eq.~(\ref{eq:ElectrFieldPasveer}) can be considered as universal because these equations do not even contain the localization length $a$. Furthermore this result shows the apparent deficiency of doing physics by computer simulations. Parameterized phenomenological equations, such as Eq.~(\ref{eq:ElectrFieldBassler}) and Eq.~(\ref{eq:ElectrFieldPasveer}), do not contain the material parameter $a$, which is decisive for the field--dependent mobility $\mu(F)$, as evidenced in Fig.~\ref{fig:Baessler_vs_Pasveer}.

Being interested in the dependence $\mu(F)$ for realistic spatially disordered systems rather than for cubic grids, we will consider in the rest of this paper a system of sites distributed in space randomly.

\section{Localization length determines $\mu(F)$ for random sites.}
\label{sec:Teff simulations}

In order to discern the decisive length scale ($\alpha$, $N^{-1/3}$, or some combination of these parameters) for the field dependence of $\mu$ in a system of random sites, we performed computer simulations using the standard Monte Carlo procedure. A disordered system is created with $140\times 140\times 140$ sites distributed randomly in a box of $L=140$, so that the average inter-site distance $N^{-1/3}$ is unity. The site energies are chosen randomly according to the DOS given in Eq.~(\ref{DOS_Gauss}). A single electron is placed onto a random site $i$ and in each simulation step performs a hopping transition to another site $j$ with probabilities weighted by the MA hopping rates given by Eqs. (\ref{eq:nu-ij}) and (\ref{eq:gamma}). After each hop, the system time is advanced by $\tau = \nu_{ij}^{-1}$. Initially, the electron is allowed to make $5\times 10^{7}$ relaxation hops to ensure steady-state conditions, after which statistics is collected for $5\times 10^{8}$ hopping transitions. The simulations were repeated and averaged $20$ times. The realistically chosen parameters were $\sigma/kT=4$ and $0.18 \leq \alpha/N^{-1/3} \leq 0.30$.

\begin{figure}
	\includegraphics[width=\linewidth]{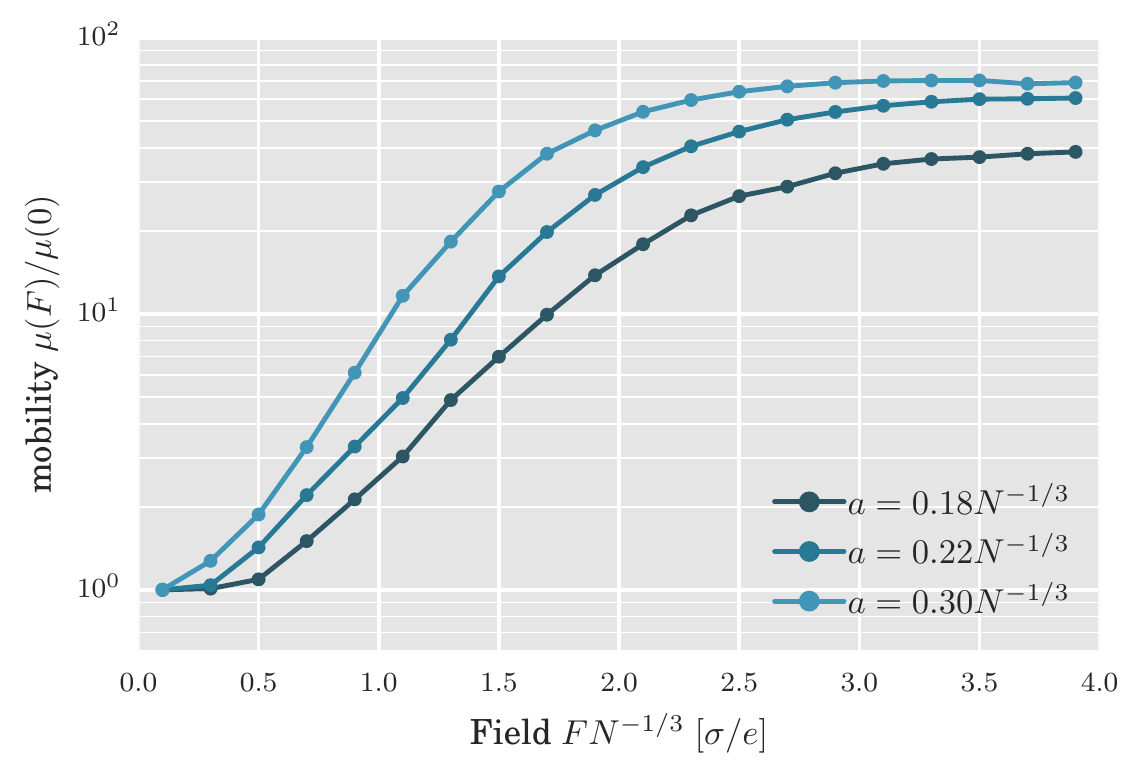}
	\caption{Normalized mobility $\mu(F)/\mu(0)$ for the system of random sites at $\sigma/kT =4$ and different values $\alpha/N^{-1/3}$ plotted vs $FN^{-1/3}/(\sigma/e)$.}
	\label{fig:nonuniversality}
\end{figure}

\begin{figure}
	\includegraphics[width=\linewidth]{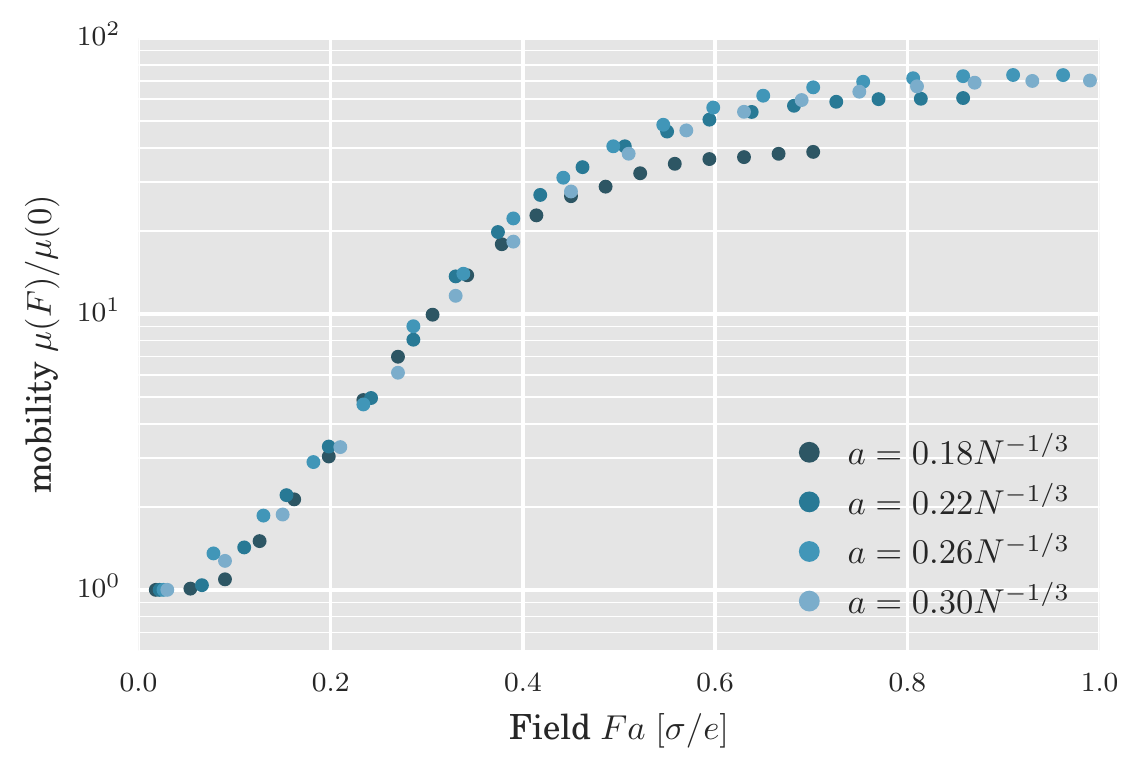}
	\caption{Normalized mobility $\mu(F)/\mu(0)$ for the system of random sites at $\sigma/kT =4$ and different values $\alpha/N^{-1/3}$ plotted vs $F\alpha/(\sigma/e)$.}
	\label{fig:universality}
\end{figure}

Our simulation results for $\mu(F)/\mu(0)$ are plotted versus $FN^{-1/3}$ in Fig.~\ref{fig:nonuniversality} and versus $F\alpha$ in Fig.~\ref{fig:universality}.
The results look really remarkable. While the plots as a function of $FN^{-1/3}$ differ from each other for different values of the parameter $\alpha/N^{-1/3}$, as they do in the case of the lattice model shown in Fig.~\ref{fig:Baessler_vs_Pasveer}, the data fall onto a universal curve when plotted as a function of $F\alpha$. The deviations for $\alpha = 0.18N^{-1/3}$ at high $F$ are caused by the effect of the negative differential conductivity discussed elsewhere\cite{Nguyen1981,Nenashev2008NDC}. The universality of plots $\mu(F)/\mu(0)$ versus $F$ in units $\sigma/e\alpha$ proves that the localization length $\alpha$, and not the intersite distance $N^{-1/3}$ (present in Eq.~(\ref{eq:ElectrFieldPasveer}) in the form of the lattice constant $b$), is the decisive length scale for the dependence $\mu(F)/\mu(0)$.

\section{Effective temperature dependent on $a$ is responsible for $\mu(F)$}
\label{sec:Teff}

One might be tempted to invent new phenomenological fitting equations in the spirit of Eqs.~(\ref{eq:ElectrFieldBassler}) and (\ref{eq:ElectrFieldPasveer}) for $\mu(F)/\mu(0)$ that would take into account the effect of the parameter  $\alpha$. Instead we suggest to recall the idea by Shklovskii, who  recognized the importance of the localization length $\alpha$ for the dependence $\mu(F)$ more than 40 years ago\cite{Shklovskii1973}.

Let us try to understand, why the localization length $\alpha$ and not the intersite distance $N^{-1/3}$ is the decisive length scale for the field dependence of $\mu(F)/\mu(0)$. Shklovskii\cite{Shklovskii1973} considered the case $T=0$ and pointed out that when a charge carrier tunnels in the field direction over some distance $x$, its energy gain due to the applied electric field amounts to $\delta = eFx$. The tunneling probability $\nu(x)\propto \exp(-2x/\alpha)$ can then be rewritten as $\nu(\delta)\propto \exp(-\delta/kT_\text{eff})$ with $T_\text{eff} \simeq eF\alpha/2$.

For the case of finite temperatures, i.e., for $T\neq 0$, Shklovskii\cite{Shklovskii1973,Shklovskii1990Fritzsche,Marianer1992} and successors\cite{Hess1993,Cleve1995,Jansson2008PRB} have shown that the combined effects of the electric field $F$ and temperature $T$ on the hopping mobility can be expressed in the form of the so-called effective temperature
\begin {equation}
\label {eq:Teff}
T_\text{eff} = \left[ T^2 + \left(\gamma
\frac{e F \alpha} {k} \right)^2 \right]^{1/2}
\end {equation}
with $\gamma \approx 0.67$.\cite{Marianer1992,Hess1993}

This result is non-trivial and it looks counterintuitive. The electric field enters the theory only via Eq.~(\ref{eq:nu-ij}), i.e., via the combination $e\mathbf{F}\cdot\mathbf{r}_{ij}$.  The length of a hop $|\mathbf{r}_{ij}|$ is of the order of the intersite distance $N^{-1/3}$. Therefore, one might expect the combination of parameters $e N^{-1/3} F$ to be essential for the field-dependent mobility. Shklovskii instead argued \cite{Shklovskii1973} that the localization length $\alpha$, i.e. the feature of a single localized state, and not the intersite distance $N^{-1/3}$ is responsible for $\mu(F)$. Taking into account that the Stark effect (determined by the length $a$) is not considered, this proposition sounds revolutionary. Only very recently it has been proven \cite{Nenashev_ARKHIV2017} that indeed $\alpha$ and not the intersite distance $N^{-1/3}$ is responsible for the dependence $\mu(F)/\mu(0)$, as described in Sec.~\ref{sec:Teff simulations}.

The counterintuitive and revolutionary nature of Shklovskii's idea might be the reason for the fact that it was ignored by the broad scientific community. For instance, in recent review papers \cite{Kuik2014,Laquai2015}, Eqs.~(\ref{eq:ElectrFieldBassler}) and (\ref{eq:ElectrFieldPasveer}) are considered as the main theoretical achievement in the study of charge transport in ODSs.  Another possible reason might be the lack of a straightforward proof for this rather counterintuitive concept.
Notably, it has never been shown before that only $\alpha$, and not the intersite distance $N^{-1/3}$, is responsible for the dependence $\mu(F)/\mu(0)$, as the parameter $N^{-1/3}/\alpha$ was always fixed and not varied in the simulations confirming the validity of Eq.~(\ref{eq:Teff}) \cite{Marianer1992,Hess1993,Cleve1995,Jansson2008PRB}. For instance, Marianer and Shklovskii \cite{Marianer1992} suggested Eq.~(\ref{eq:Teff}) as the result of computer simulations using the fixed value $N^{-1/3}/a =3$. Their result can be plotted as a function of $eaF$, and, with the same success, as a function of $eN^{-1/3}F/3$. The data in Sec.~\ref{sec:Teff simulations} and in the previous paper \cite{Nenashev_ARKHIV2017} prove, however, that the localization length $\alpha$, and not the intersite distance $N^{-1/3}$ is responsible for the dependence $\mu(F)/\mu(0)$, as assumed in the concept of the effective temperature.

\begin{figure}
	\includegraphics[width=\linewidth]{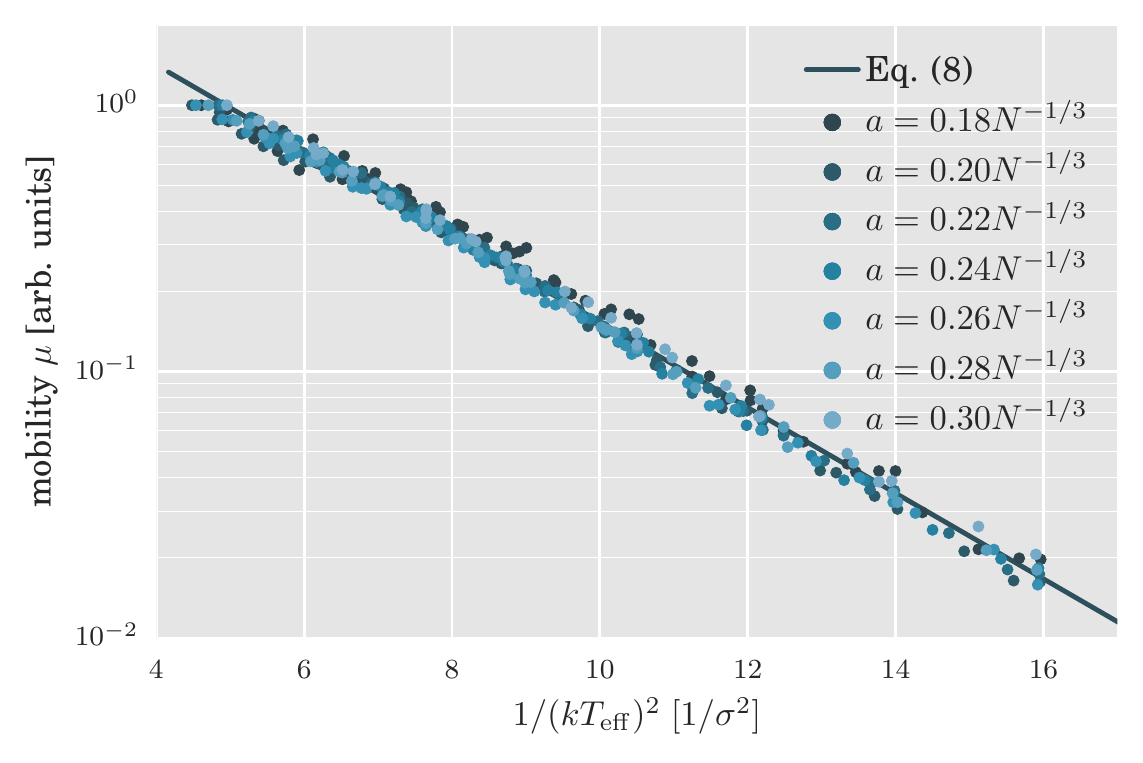}
	\caption{Dependence of the mobility on $1/(kT_{\text{eff}})^{2}$ for the system of random sites.}
	\label{fig:T_eff}
\end{figure}

Following this concept \cite{Shklovskii1973,Shklovskii1990Fritzsche,Marianer1992,Hess1993,Cleve1995,Jansson2008PRB}, the dependence of the charge carrier mobility $\mu(F)$ on the electric field in hopping conduction can be obtained by inserting the effective temperature $T_{\text{eff}}$ on the place of the laboratory temperature $T$ in the analytical expressions for $\mu(T)$  obtained at low $F$. The temperature dependence of hopping mobility $\mu(T)$ in the GDM at low carrier concentrations is known to have the form\cite{Bassler1993}
\begin{align}
	\mu \propto \exp \left[
	- C\left(\frac{\sigma}{kT}\right)^{2}\right] \, ,
	\label{eq:mu-T}
\end{align}
where the coefficient $C$ has typically the value $C\approx 0.4$, only slightly depending on the ratio $N^{-1/3}/\alpha$.\cite{Baranovskii2000} In Fig.~\ref{fig:T_eff}, the mobility $\mu$, obtained in our simulations is  
plotted as a function of $(\sigma/kT_{\text{eff}})^{2}$, where $T_{\text{eff}}$ is given by  Eq.~(\ref{eq:Teff}) with $\gamma = 0.67$. The results perfectly agree with the prediction of Eq.~(\ref{eq:mu-T}) with $T = T_{\text{eff}}$, $C = 0.37$, as shown in Fig.~\ref{fig:T_eff} by the solid line. Simulations for Fig.~\ref{fig:T_eff} were carried out for the parameters sets $\sigma/kT$ between 3 and 4 with the step size 0.25 and $eFN^{-1/3}/\sigma$ between 0.1 and 3.9 with the step size 0.2. The values of $\mu$ in Fig.~\ref{fig:T_eff} are normalized by the mobility values at highest $F$ and $T$.

Experimental data for the field-dependent mobility at low carrier concentrations should be compared not with Eq.~(\ref{eq:ElectrFieldBassler}), or Eq.~(\ref{eq:ElectrFieldPasveer}), but rather with Eq. (\ref{eq:mu-T}), in which temperature $T$ is replaced by the field-dependent effective temperature $T_{\text{eff}}$ given by Eq.~(\ref{eq:Teff}).  Such a comparison allows one to determine the value of the localization length $\alpha$ experimentally.

At high carrier concentrations $n$, the temperature dependence of the mobility is described \cite{Baranovskii2002a,Baranovski2006,Baranovskii2014,Nenashev_Topical_2015} by the Arrhenius law instead of Eq. (\ref{eq:mu-T}). In order to describe the dependence $\mu(F)$ in this regime, one should replace the temperature $T$ in the Arrhenius equation with the effective temperature given by Eq.~(\ref{eq:Teff}).

\section{Effective temperature proven analytically}
\label{sec:Teff analytical}

In Sec. \ref{sec:Teff}, we provided analytical arguments for the validity of the effective--temperature concept, where $T_{\text{eff}}$ only depends on the localization length $a$, as suggested by Shklovskii \cite{Shklovskii1973} at $T=0$. Below we provide additional arguments \textit{valid also at finite }$T$ in favor of the localization length $\alpha$ as the decisive spatial scale responsible for $\mu(F)/\mu(0)$ in a hopping motion of charge carriers via random sites. The results proven in the rest of this report can be formulated as follows.

i) The energy distribution of charge carriers does not depend on the concentration of sites $N$ at a fixed ratio $n/N$.

ii) At non-zero field and temperature the distribution function is the Fermi function with the effective temperature, independent of the site concentration $N$.

The applied electric field changes the rates of carrier transitions between sites, so that site occupation probabilities at nonzero field can differ from their equilibrium values given by the Fermi--Dirac distribution. Here, we will show that one can find the occupation probabilities in the presence of an electric field from a simple integral equation, Eq.~(\ref{eq:balance}), \emph{assuming that these probabilities are the same for all sites of the same energy}, as is granted for zero field. Herewith the occupation probability of some site $i$ is a function $f(\ve_i)$ solely of the site energy $\ve_i$, as in the case at zero field. Note that the energy of a site is considered \emph{without} the contribution of the potential of the external field. The sites are considered as randomly placed in space, without correlations between their positions and energies. No further assumptions will be involved.

Consider all the carrier transitions from sites within some energy range $[\ve_1, \, \ve_1+\mathrm{d}\ve_1]$ to sites within another range $[\ve_2, \, \ve_2+\mathrm{d}\ve_2]$, where widths $\mathrm{d}\ve_1$ and $\mathrm{d}\ve_2$ are small compared to $kT$. Let us denote as $\mathcal{R}(\ve_1,\ve_2) \, \mathrm{d}\ve_1 \, \mathrm{d}\ve_2$ the number of such transitions per unit time in the whole sample. Then,
\begin{multline}
	\label{eq:rate-from-to}
	\mathcal{R}(\ve_1,\ve_2) \, \mathrm{d}\ve_1 \, \mathrm{d}\ve_2 = \\
	\sum\limits_{\substack{i \\ \ve_i \in [\ve_1, \ve_1+\mathrm{d}\ve_1]}} \quad
	\sum\limits_{\substack{j \\ \ve_j \in [\ve_2, \ve_2+\mathrm{d}\ve_2]}}
	f(\ve_i) \, [1-f(\ve_j)] \, \nu_{ij} \, .
\end{multline}

Since site positions and energies are uncorrelated, the vectors $\mathbf{r}_{ij}$ are uniformly distributed over the three-dimensional vector space with the density $V \rho(\ve_1) \, \mathrm{d}\ve_1 \, \rho(\ve_2) \, \mathrm{d}\ve_2$, where $V$ is the volume of the sample. If $V$ is large enough, the vectors $\mathbf{r}_{ij}$ fill the space densely enough to enable integration instead of summation in Eq.~(\ref{eq:rate-from-to}),
\begin{equation}
	\label{eq:from-sum-to-int}
	\sum_i \sum_j \quad \Rightarrow \quad
	V \rho(\ve_1) \, \mathrm{d}\ve_1 \, \rho(\ve_2) \, \mathrm{d}\ve_2
	\int\limits_0^\infty r^2 \mathrm{d}r \int\limits_0^\pi 2\pi\sin\theta \, \mathrm{d}\theta ,
\end{equation}
where polar coordinates $r$ and $\theta$ are introduced in the space of vectors $\mathbf{r}_{ij}$. Directing the polar axis along the field~$\mathbf{F}$, and taking into account that $\ve_i=\ve_1$ and $\ve_j=\ve_2$ to the accuracy of $\mathrm{d}\ve_1$ and $\mathrm{d}\ve_2$, one obtains from Eqs.~(\ref{eq:rate-from-to})--(\ref{eq:from-sum-to-int}) the following representation for the rate $\mathcal{R}(\ve_1,\ve_2)$,
\begin{equation}
	\label{eq:rate-from-to-2}
	\mathcal{R}(\ve_1,\ve_2) = V \rho(\ve_1) \, \rho(\ve_2) \, f(\ve_1) \, [1-f(\ve_2)] \, \mathcal{F}(\ve_2-\ve_1) ,
\end{equation}
where
\begin{multline}
	\label{eq:F-function}
	\mathcal{F}(\Delta\ve) = 2\pi\nu_0  \\
	\times \int_0^\infty e^{-2r/\alpha} \left[ \int_0^\pi \gamma(\Delta\ve+eFr\cos\theta) \sin\theta \, \mathrm{d}\theta \right] r^2 \mathrm{d}r .
\end{multline}

Now it becomes easy to formulate the carrier balance equation in the steady state. The rate of carrier transitions from the vicinity of energy $\ve_1$ to the vicinity of $\ve_2$ is proportional to $\mathcal{R}(\ve_1,\ve_2)$, and the rate of reverse transitions is proportional to $\mathcal{R}(\ve_2,\ve_1)$. Integration of these rates over~$\ve_2$ provides the total carrier loss from/gain to the energy~$\ve_1$, and the equality of loss and gain determining the steady state,  takes the form
\begin{equation}
	\label{eq:balance-R}
	\int\limits_{-\infty}^{+\infty} \mathcal{R}(\ve_1,\ve_2) \, \mathrm{d}\ve_2 = \int\limits_{-\infty}^{+\infty} \mathcal{R}(\ve_2,\ve_1) \, \mathrm{d}\ve_2 \, .
\end{equation}
Inserting Eq.~(\ref{eq:rate-from-to-2}), one obtains the following balance equation:
\begin{multline}
	\label{eq:balance}
	f(\ve_1) \int\limits_{-\infty}^{+\infty} \rho(\ve_2) \, [1-f(\ve_2)] \, \mathcal{F}(\ve_2-\ve_1) \, \mathrm{d}\ve_2 =  \\
	[1-f(\ve_1)] \int\limits_{-\infty}^{+\infty} \rho(\ve_2) \, f(\ve_2) \, \mathcal{F}(\ve_1-\ve_2) \, \mathrm{d}\ve_2
\end{multline}

This is the master equation for calculating the carrier distribution function $f(\ve)$ in the presence of the external electric field. It proves that $f(\ve)$ does not depend on the site concentration $N$. Indeed, since $N$ contributes to this equation only as a factor in the density of states $\rho(\ve_2)$, it is present in both sides of the equation, and the factors $N$ cancel. Therefore, the electric field $F$ affects the carrier energy distribution only in the combination $eF\alpha$, but not in the combination $eFN^{-1/3}$. This supports our data obtained by Monte Carlo simulations depicted in Fig.~\ref{fig:universality}.

The question might arise on how sensitive this conclusion is with respect to the choice of the expression for transition rates. So far we considered the Miller-Abrahams expression given by Eqs.~(\ref{eq:nu-ij}), (\ref{eq:gamma}). We would like to emphasize that the electric field $F$ affects the carrier energy distribution only in the combination $eF\alpha$ also for all other shapes of the transition rates, in which the distance $r$ of a hop appears in the combination $r/\alpha$. If the transition rate can be represented in the form of Eq.~(\ref{eq:nu-ij}), Eqs.~(\ref{eq:rate-from-to})-(\ref{eq:balance}) keep their form. Therefore the conclusion about the decisive role of the localization length $\alpha$ as the only relevant length scale for the dependence $\mu(F)/\mu(0)$ is valid also for Marcus transition rates, which have the form of Eq.~(\ref{eq:nu-ij}), though depending on the matrix renormalization energy\cite{Marcus1964}.

Furthermore, Eq.~(\ref{eq:balance}) provides a basis for justifying the concept of the effective temperature \emph{at nonzero temperatures $T$}. To see this, let us first note that at zero electric field, the following relation holds for any $\Delta\ve$:
\begin{equation}
	\label{eq:ratio-F-no-field}
	\frac {\mathcal{F}(\Delta\ve)} {\mathcal{F}(-\Delta\ve)} = \exp\left( -\frac{\Delta\ve}{kT} \right)  \qquad \text{(at $F=0$)} ,
\end{equation}
as evident from Eqs.~(\ref{eq:gamma}) and~(\ref{eq:F-function}). In this case, according to detailed balance, the solution of the master equation Eq.~(\ref{eq:balance}) is the Fermi--Dirac distribution. Similarly, if (at nonzero electric field) there is such a quantity $T_{\text{eff}}$ that for any $\Delta\ve$
\begin{equation}
	\label{eq:ratio-F-Teff}
	\frac {\mathcal{F}(\Delta\ve)} {\mathcal{F}(-\Delta\ve)} \approx \exp\left( -\frac{\Delta\ve}{kT_{\text{eff}}} \right) ,
\end{equation}
then the solution $f(\ve)$ of Eq.~(\ref{eq:balance}) should have the form of the Fermi--Dirac function with the effective temperature $T_{\text{eff}}$ instead of the real temperature $T$:
\begin{equation}
	\label{eq:Fermi-Teff}
	f(\ve) \approx \left[ \exp\left( \frac{\ve-\ve_f}{kT_{\text{eff}}} \right) + 1 \right]^{-1} ,
\end{equation}
with an appropriate value of the Fermi energy $\ve_f$.

Using Eq.~(\ref{eq:F-function}), we numerically checked the validity of the relation in Eq.~(\ref{eq:ratio-F-Teff}) for the whole range of electric fields at $\sigma/kT = 3$ and  $\sigma/kT = 4$. This relation is proven to hold in the range of energy differences $\Delta\ve \in [-\sigma^2/kT, \; \sigma^2/kT]$, which has been proven\cite{Bassler1993,Baranovski2006,Tessler2009,Baranovskii2014,Nenashev_Topical_2015,Ostroverkhova2016} as decisive for charge transport in organic semiconductors. Moreover, in the limit of small carrier concentrations we verified that the solution $f(\ve)$ of Eq.~(\ref{eq:balance}) follows Eq.~(\ref{eq:Fermi-Teff}) in the important energy range $\ve \in [-\sigma^2/kT, \; 0]$. Herewith it is apparent from the above consideration that the effective temperature introduced in Eqs.~(\ref{eq:ratio-F-Teff}) and (\ref{eq:Fermi-Teff}) cannot depend on the site concentration $N$, so that the electric field contributes to the effective temperature only in the combination $eF\alpha$.

\section{Conclusions}
\label{sec:conclusions}

By computer simulations and by analytical calculations we showed that the localization length $a$ of charge carriers in the localized states is the \textit{only} spatial parameter responsible for the dependence of the hopping mobility $\mu$ on the applied electric field $F$ in a system of random sites. Remarkably, this parameter $\alpha$ is not present in Eqs.~(\ref{eq:ElectrFieldBassler}) and (\ref{eq:ElectrFieldPasveer}), which are often treated as theoretical predictions for $\mu(F)$ and used in device simulations.  Results of the current report exclude Eqs.~(\ref{eq:ElectrFieldBassler}) and (\ref{eq:ElectrFieldPasveer}) as candidates for the description of the dependence $\mu(F)/\mu(0)$. Instead, the effective temperature that contains the electric field $F$ in the combination $eF\alpha$ is responsible for the dependencies of the carrier mobility $\mu$ on $T$ and $F$, as illustrated in Fig.~\ref{fig:T_eff}.

In essence, theories developed for the temperature--dependent hopping mobility $\mu(T)$ at vanishingly small electric fields $F$, as described in recent reviews \cite{Baranovskii2014,Nenashev_Topical_2015}, are capable to account also for the field--dependent mobility at high $F$ if the temperature $T$ in the low--field theories is replaced by the field--dependent effective temperature $T_{\text{eff}}(F,T)$ given by Eq.~(\ref{eq:Teff}).

\begin{acknowledgments}
Authors are indebted to Boris Shklovskii for valuable comments.	Financial support of the Deutsche Forschungsgemeinschaft (GRK 1782) is gratefully acknowledged.
\end{acknowledgments}

\bibliography{article}

\end{document}